\documentclass{nature}
\usepackage{latexsym}
\usepackage{amsthm}
\usepackage{amssymb}
\usepackage{amsmath}
\usepackage[all]{xy}
\usepackage{graphicx}
\usepackage{subfigure}
\usepackage{dcolumn}
\usepackage{bm}
\usepackage{bbm}
\usepackage{dsfont}
\usepackage{color}
\usepackage{bbold}
\usepackage[usenames,dvipsnames]{xcolor}
\usepackage{amsfonts}
\usepackage{cases}
\usepackage[a4paper,colorlinks=true,
linkcolor=blue,citecolor=blue,
pdfauthor={ },
pdftitle={ },
pdfsubject={ },
pdfkeywords={ }]{hyperref}
\usepackage{epstopdf}
\newcommand{\onlinecite}[1]{\hspace{-1 ex} \nocite{#1}\citenum{#1}}
\makeatletter

\newcommand{\Rmnum}[1]{\expandafter\@slowromancap\romannumeral #1@}
\makeatother
\begin{document}

\title{Experimental Benchmarking of Quantum Control in Zero-Field Nuclear Magnetic Resonance}

\author{Min Jiang$^{1,2}$$^*$, Teng Wu$^{2,3}$$\thanks{These authors contributed equally to this work.}$ \thanks{e-mail: teng@uni-mainz.de}, John W. Blanchard$^{3}$\thanks{e-mail: blanchard@uni-mainz.de}, Guanru Feng$^4$, Xinhua Peng$^1$ \thanks{e-mail: xhpeng@ustc.edu.cn}, \mbox{and Dmitry Budker$^{2,3,5}$}}
\maketitle
\begin{affiliations}
\item Hefei National Laboratory for Physical Sciences at the Microscale and Department of Modern Physics, University of Science and Technology of China, Hefei, 230026, China
\item Johannes Gutenberg-University Mainz, 55128 Mainz, Germany
\item Helmholtz-Institut Mainz, 55099 Mainz, Germany
\item Institute for Quantum Computing and Department of Physics and Astronomy, University of Waterloo, Waterloo, Ontario, N2L 3G1, Canada
\item Department of Physics, University of California at Berkeley, California 94720-7300, USA
\end{affiliations}


\begin{abstract}
Zero-field nuclear magnetic resonance (NMR) provides complementary analysis modalities to those of high-field NMR and allows for
ultra-high-resolution spectroscopy and measurement of untruncated spin-spin interactions.
Unlike for the high-field case, however, universal quantum control -- the ability to perform arbitrary unitary operations -- has not been experimentally demonstrated in zero-field NMR.
This is because the Larmor frequency for all spins is identically zero at zero field, making it challenging to individually address different spin species.
We realize a composite-pulse technique for arbitrary independent rotations of $^1$H and $^{13}$C spins in a two-spin system.
Quantum-information-inspired randomized benchmarking and state tomography are used to evaluate the quality of the control.
We experimentally demonstrate single-spin control for $^{13}$C with an average gate fidelity of $0.9960(2)$ and two-spin control via a controlled-not (CNOT) gate with an estimated fidelity of $0.99$.
The combination of arbitrary single-spin gates and a CNOT gate is sufficient for universal quantum control of the nuclear spin system.
The realization of complete spin control in zero-field NMR is an essential step towards applications to quantum simulation, entangled-state-assisted quantum metrology, and zero-field NMR spectroscopy.

\end{abstract}


Zero-field nuclear magnetic resonance (NMR) is an alternative magnetic resonance modality where nuclear-spin information is measured in the absence of applied magnetic field\cite{Weitekamp1983, Lee1987, LedbetterMP2008, TheisT2011, BlanchardD2016},
and serves as a complementary analysis tool to conventional high-field NMR.
Zero-field NMR experiments regularly achieve nuclear spin coherence times longer than ten seconds\cite{Ledbetter2012, Blanchard2013, Emondts2014} without using dynamical-decoupling pulse sequences.
Whereas at high field the spins are coupled more strongly to the magnetic field than to each other, at zero field the spins are strongly coupled to each other by spin-spin couplings.
Significantly, at zero field, the spin-spin couplings are not truncated as they are in conventional NMR.
This means that zero-field NMR is capable of measuring certain spin-dependent interactions\cite{Blanchard2015, King2017},
which are not generally accessible in high-field NMR experiments.

High-fidelity control of nuclear spins is important for NMR applications,
ranging from coherent spectroscopy\cite{Ernst1987, Glaser1998} to quantum information processing (QIP)\cite{Gershenfeld1997, Nielsen2000, Bennett2000}.
In fact, QIP and magnetic resonance face a number of common issues in the control of spin dynamics and optimization of signals.
On the one hand, magnetic resonance can provide a physical platform for QIP\cite{Jones2000, Vandersypen2001}.
On the other hand, precise nuclear-spin control methods developed for QIP are also valuable for NMR signal enhancement and pulse sequence design\cite{Jenista2009, Chang2015, Bonnard2012, Nimbalkar2013, Holbach2014, Holbach2015}.
There are two basic experimental approaches to evaluating the quality of the control in QIP.
One is quantum process tomography (QPT)\cite{Vandersypen2005} allowing for complete characterization of control operations,
the other is randomized benchmarking (RB)\cite{PRA_RBQG, Magesan2011}, which reveals average gate errors.
These have been employed in various systems, e.g., trapped ions\cite{Riebe2006, Gaebler2012}, NMR\cite{Childs2001, Ryan2009}, and superconducting circuits\cite{Chow2009, Barends2014}.

In this work, we experimentally demonstrate and quantify the fidelity of universal quantum control of a spin system composed of two coupled heteronuclear spins at zero magnetic field.
Spins with different gyromagnetic ratios all have identical (zero) Larmor frequency at zero field, and thus individual manipulation of the different spin species presents a challenge.
To overcome this challenge, we use a composite sequence to rotate one of the two spins by a desired angle and to cancel the accumulated rotation angle of the other\cite{Bian2017}.
Based on this,
we realize single-spin gates for $^{13}$C and $^1$H and a two-spin controlled-not (CNOT) gate in $^{13}$C-formic acid ($^1$H-$^{13}$COOH, where the acidic proton is neglected due to rapid exchange).
Randomized benchmarking is implemented to estimate the single-spin gate fidelity for $^{13}$C to be $0.9960(2)$.
Utilizing the single-spin gates, a temporal averaging technique\cite{Knill1998pure} is developed to realize state tomography at zero field for determining the quantum state of nuclear spins.
By performing state tomography before and after application of the CNOT gate, we are able to characterize the performance of the CNOT gate
using a constrained fitting technique, which yields an estimated gate fidelity of $0.99$.
We also evaluate the nature of the dominant errors for nuclear-spin control in zero-field NMR.

\section*{Results}
\subsection{Spin system at zero magnetic field.}
\begin{figure}[http]
\centering
\includegraphics[width=1\columnwidth]{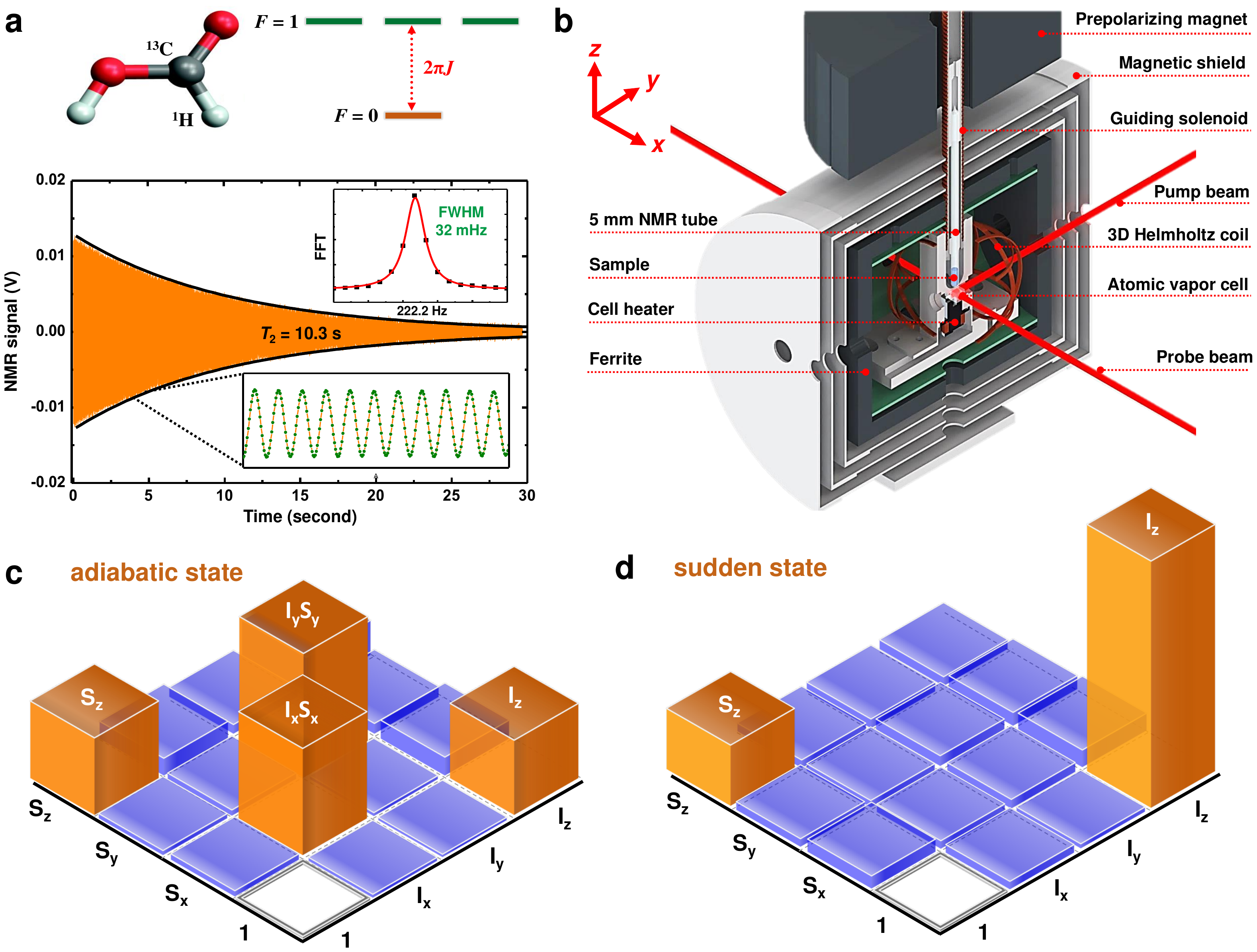}
    \caption{$\textbf{a}$, Schematic atomic structure and energy levels of ${}^{13}$C-formic acid ($^1$H-$^{13}$COOH); single-shot zero-field NMR signal. At zero field, the eigenstates of $^{13}$C-$^{1}$H group are triplet states ($F=1$, $|T_{-1,0,+1}\rangle$) and singlet state ($F=0$, $|S_0\rangle$), where the subscripts indicate the magnetic quantum numbers. The decay time of the signal is measured to be $T_2 = 10.3$~s. The single peak of the fast Fourier transform signal is related to coherence between $|T_0\rangle$ and $|S_0\rangle$, which produces $z$-magnetization oscillating with frequency $J=222.2176(1)$~Hz. The FWHM (full width at half maximum) linewidth obtained from a Lorentzian fit is $32$~mHz. $\textbf{b}$, Experimental setup for zero-field NMR spectroscopy, described in $\textbf{Methods}$. The NMR sample is contained in a $5$-mm NMR tube, and pneumatically shuttled between a $1.8$~T prepolarizing magnet and the interior of a four-layer magnetic shield. A guiding field is applied in the $z$ direction during the pneumatic shuttling. NMR signals are detected with an atomic magnetometer with a $^{87}$Rb vapor cell operating at $180$~$^\circ$C.  $\textbf{c}$ and $\textbf{d}$, Results of state tomography on initial states after adiabatic (c) and sudden (d) transfers.
 }
    \label{Fig1}
\end{figure}

A liquid-state $n$-spin system at zero magnetic field can be described by the Hamiltonian:
$
H_J  = \sum\limits_{i ; j>i}^n {{2 \pi J_{ij}}}  \mathbf{I}_i \cdot \mathbf{I}_j
$
, where ${J_{ij}}$ is the strength of the scalar spin-spin coupling between the $i$th and $j$th spins,
$\mathbf{I}_i = (I_{ix}, I_{iy}, I_{iz})$ is the spin angular momentum operator of the $i$th spin,
and the reduced Planck constant is set to one.
We experimentally demonstrate feasibility of our nuclear-spin control scheme by using $^{13}$C-formic acid (Fig. \ref{Fig1}a), a convenient heteronuclear two-spin system.
At zero magnetic field, the eigenstates of a two-spin-$1/2$ system are most conveniently defined in terms of the total angular momentum $\mathbf{F}=\mathbf{I}_1+\mathbf{I}_2$, yielding a singlet state with $F=0$ and three degenerate triplet states with $F=1$ (Supplementary Section \Rmnum{1}).
The nuclear-spin singlet state is antisymmetric with respect to exchange,
and it cannot evolve into symmetric states under the symmetric intramolecular dipole-dipole interaction\cite{Carravetta2004}.
For this reason, the lifetime of a nuclear-spin singlet state can be long.
The singlet state lifetime of the $^{13}$C-formic acid sample used in our experiment is measured to be $16.7$~s.
While homonuclear singlets are long-lived at arbitrary fields, this holds for heteronuclear systems only in near-zero fields\cite{Emondts2014}, where the lifetime of the singlet-triplet coherence can also be enhanced\cite{Sarkar2010}.
The lifetime of the singlet-triplet coherence observed in our experiment is $T_2= 10.3$~s, as shown in Fig. \ref{Fig1}a.
A long coherence time is important for nuclear-spin control, as numerous coherent operations can be implemented (e.g., nearly $10^{5}$ single-spin gates or $10^{4}$ two-spin gates in this experiment).
This is also useful for molecular structure determination and fundamental physics,
as it permits high resolution of minute frequency differences and precise determination of long-range spin-spin interactions.

Experiments are performed using an apparatus similar to that of Refs~\onlinecite{Ledbetter2009, TheisT2011, TaylerMC2017} which is schematically shown in Fig. \ref{Fig1}b.
The sample ($\sim 200$~$\mu$L) contained in a $5$-mm NMR tube is polarized in a permanent magnet,
and then pneumatically shuttled through a guiding solenoid to a zero-field region.
The bottom of the NMR tube is at a distance of $\sim1$~mm above a rubidium vapor cell of an atomic magnetometer\cite{Allred2002, Budker2007}.
Details of the experimental setup are described in $\textbf{Methods}$.
A guiding magnetic field ($\sim 3\times 10^{-5}$~T) is applied along the transfer direction of the sample during the transfer,
and is turned off after the sample is transferred to the zero-field region.
The way that the guiding field is switched to zero plays a crucial role in determining the initial state.
There are two limiting cases, which correspond to sudden and adiabatic changes.
For brevity, we call the resulting spin states ``sudden'' and ``adiabatic'' states, respectively (see $\textbf{Methods}$).
In order to characterize the initial state, state tomography is performed,
which helps us optimize experimental parameters to make the initial state close to one of these two limiting cases.
State tomography is based on the temporal averaging technique such as that implemented in Ref.~\onlinecite{Knill1998pure}.
Details are given in the Supplementary Section \Rmnum{2}.
Figures \ref{Fig1}c and d show the results for the optimized adiabatic and sudden states,
which are displayed in the Pauli basis of a two-spin system (for details of the Pauli basis, see Supplementary Section \Rmnum{2}).
The fidelities are calculated to be $F=0.98$ for the adiabatic state and $F = 0.99$ for the sudden state.

\subsection{Single-spin control.}
\begin{figure}[http]
\centering
\includegraphics[width=1\columnwidth]{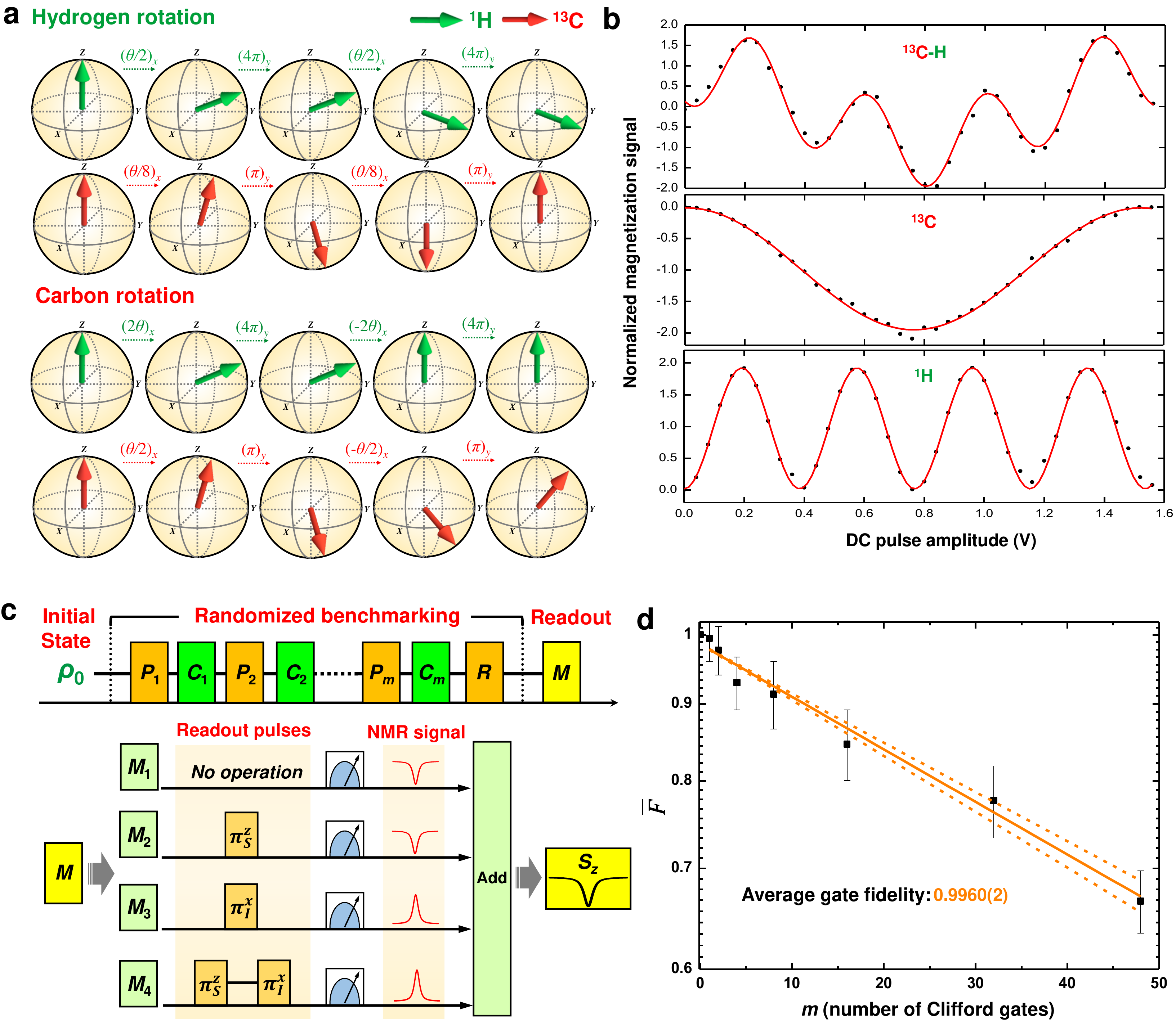}
    \caption{$\textbf{a}$, Schematic diagram of individual spin rotation for $^{1}$H (top panel) and $^{13}$C (bottom panel), as presented in text. The initial states of $^1$H and $^{13}$C are aligned to $|$$\uparrow \rangle$ for simplicity. $\textbf{b}$, Experimental verification of collective (top panel) and individual nuclear spin rotation for $^{13}$C (middle panel) and $^1$H (bottom panel). Each data point corresponds to a single measurement. Theoretical fits are shown with solid lines. $\textbf{c}$, Clifford-based randomized benchmarking. The initial state $\rho_0$ is chosen as the sudden state. Random sequences with $P =  {e}^{\pm i \pi V}$ and $C = {e}^{\pm  i \frac{\pi}{2} Q}$ are applied for each sequence length $m$, where the Clifford gates are realized by $PC$. Here, $ V \in \{ \mathds{1}, S_x, S_y, S_z \}$ and $Q \in \{ S_x, S_y, S_z \} $, where $S_x, S_y, S_z$ are spin angular momentum operators. The
recovery gate $R$ is chosen to return the system to the initial state. A set of temporal averaging sequences are used to measure $\langle S_z \rangle$.  $\textbf{d}$, Semi-log graph of $^{13}$C single-spin randomized benchmarking experimental results. Each point is an average over $32$ random sequences of $m$ Clifford gates, and the error bars indicate the standard error of the mean. A single exponential decay shown with a solid line is used to fit the fidelity decay and reveals an average gate fidelity of $0.9960(2)$.
 }
    \label{Fig2}
\end{figure}

Spin-selective coherent control of nuclear spins in high-field NMR is generally accomplished by radio-frequency pulses at the spins' resonant frequencies.
However, nuclear spins are not selectively addressable when their Larmor frequencies all become zero at zero magnetic field.
For two spins at zero field, e.g., $^{13}$C ($S$) and $^{1}$H ($I$),
the available external controls of nuclear spins are DC magnetic-field pulses along $x$, $y$, and $z$ with the Hamiltonians $H_\eta =-B_\eta (\gamma_I I_\eta + \gamma_S S_\eta), \eta =x,y,z$.
Here, $\gamma_I$ and $\gamma_S$ are the gyromagnetic ratios of the respective spins.
For ${}^{1}$H and ${}^{13}$C, the gyromagnetic ratios allow one to manipulate the spin $S$ by a $\pi$ pulse while leaving the spin $I$ effectively unchanged,
i.e., $\mathcal{U}_{\eta} ^S (\pi)= e^{-iS_{\eta} \pi} \approx e^{ -i I_{\eta} 4 \pi - i S_{\eta} \pi}$ because $\gamma_I / \gamma_S \approx 4$.
An arbitrary rotation of one of the spins, for example, $I$, can be realized by a pulse sequence that begins by rotating the spin $I$ by half of the desired angle, as shown in the top panel of Fig. \ref{Fig2}a.
This also rotates the spin $S$ by some angle.
Next, a $\pi$ pulse is applied to the spin $S$,
and then the second-half rotation is applied to the spin $I$,
followed by a $\pi$ pulse on the spin $S$.
With this sequence, the phases accumulated by the spin $S$ in the two halves of the rotation cancel.
This is also valid for arbitrary rotations of the spin $S$, except that the second half rotation is along the opposite direction, as shown in the bottom panel of Fig. \ref{Fig2}a.
A detailed description of how to implement single-spin gates is included in the Supplementary Section \Rmnum{3}.
Notably, this approach can be extended to heteronuclear multi-spin systems\cite{Bian2017}.
As discussed above, the key is to implement a $\pi$ rotation on one local spin.
The different gyromagnetic ratios for heteronuclear spins allow one to perform an odd number of $\pi$ rotations on the target nuclear spin and perform an even number of $\pi$ rotations on other spins.

We experimentally realize arbitrary individual spin rotations for $^{13}$C and $^1$H in $^{13}$C-formic acid, as shown in Fig.~\ref{Fig2}b.
The amplitude of DC pulse is calibrated by experiments similar to that of Ref.~\onlinecite{Emondts2014}.
The $\pi$ pulse on $S$, i.e., $^{13}$C, is $\sim 50\ \mu$s long.
The $^{13}$C-$^1$H nuclear-spin system is initially prepared in the adiabatic state.
A DC pulse along $x$ with amplitude $B_{\textrm{dc}}$ results in the amplitude of $z$ magnetization\cite{Emondts2014} proportional to ($\textrm{cos}\theta_S-\textrm{cos}\theta_I$), where $\theta_{S,I}=\gamma_{S,I} B_{\textrm{dc}} \tau$.
The dependence of the magnetization signal amplitude on the DC pulse amplitude along $x$ is shown in the top panel of Fig. \ref{Fig2}b.
An individual rotation of $^{13}$C results in $z$ magnetization proportional to $(\textrm{cos}\theta_S-1)$.
As previously discussed, individual rotation of $^{13}$C means that rotation is performed only on $^{13}$C spins while doing nothing (an identity operation) on $^1$H spins.
Similarly, for $^1$H, the $z$ magnetization is proportional to $(1-\textrm{cos}\theta_I)$.
The evolution under the corresponding selective pulse sequence for $^{13}$C and $^1$H is shown in the middle and the bottom panel of Fig. \ref{Fig2}b, respectively.
Our results (Fig. 2b) are in good agreement with the theoretical analysis and provide experimental parameters for realizing arbitrary single-spin gates.

To estimate the average single-qubit gate fidelity\cite{PLA_SFAGFQDO},
we adopt the Clifford-based randomized benchmarking (RB)\cite{PRA_RBQG, Ryan2009, Magesan2011} method.
Measuring the decay of the $\langle S_z \rangle$ amplitude with respect to the number ($m$) of randomized Clifford gates in the benchmarking sequence yields the average $^{13}$C single-qubit gate fidelity.
By averaging the amplitude of $\langle S_z \rangle$ over $k$ different randomized benchmarking sequences with the same length $m$, and normalizing this averaged value to that of $m=0$, the normalized signal, $\overline{F}$, can be written as
$
\overline{F}= (1-d_{if})(1-2\epsilon_g)^{m},
$
where $d_{if}$ is due to the imperfection of the state initialization and readout and $\epsilon_g$ is the average error per Clifford gate\cite{PRA_RBQG, Magesan2011}.
The average gate fidelity ($F_{\textrm{avg}} = 1 -\epsilon_g $) derived from this method is resilient to the state-preparation and measurement errors.

The randomized benchmarking pulse sequences are shown in Fig. \ref{Fig2}c.
The sudden state is selected as the initial state, as it only contains two components in the Pauli basis, $I_z$ and $S_z$.
To measure $S_z$ independently, we adopt a temporal averaging technique (see Fig. \ref{Fig2}c) where signals acquired using four different independent readout operations, $M \in \{\textrm{No operation}, \pi_S^z, \pi_I^x, \pi_S^z\textrm{-}\pi_I^x\}$, are averaged together (further details in the Supplementary Section \Rmnum{2}).
We generate $k=32$ random sequences for each $m$.
As shown in Fig. \ref{Fig2}d, the randomized benchmarking results yield an average error per Clifford gate $\epsilon_g=0.0040(2)$ and an imperfection of the state initialization and readout $d_{if}=0.0141$.
The average $^{13}$C single-spin gate fidelity is $F_{\textrm{avg}}=1-0.0040(2)=0.9960(2)$.

In general, errors in the control of quantum systems can be classified into three categories, i.e., unitary, decoherent, and incoherent errors\cite{Pravia2003}.
For our experiment, it is the unitary error which comes from pulse imperfections (amplitude miscalibration and direction misalignment) that principally limits the single-spin gate fidelity.
The decoherent error can be neglected due to the fact that the coherence time of our system is sufficiently longer than the entire duration of the sequence.
The incoherent error, which mainly comes from pulse-field inhomogeneity measured to be $\sim 0.2\%$ (see Supplementary Figure S$2$) over the sample volume,
is estimated to be about $10^{-5}$ per gate and is as well much smaller than the experimentally measured average gate error.


\subsection{Two-spin CNOT gate.}
The conventional way to generate a CNOT gate is to utilize the $I_z S_z$ (Ising) interaction combined with single-spin operations\cite{Vandersypen2005}.
However, at zero magnetic field, the scalar spin-spin coupling retains $I_xS_x$, $I_yS_y$, and $I_zS_z$ terms.
An effective $I_zS_z$ interaction can be realized by implementing a pulse sequence in which a $z$-$\pi$ pulse is at first applied to the spin $S$, the system is allowed to evolve for a time $t_p/2$, 
followed by another $z$-$\pi$ pulse of opposite sign, and a second $t_p/2$ free evolution period.
This process can be expressed as the following propagator:
$
 \mathcal{U}_{zz} (\theta) = e^{-i H_{J} t_p/2}  \mathcal{U}_{z} ^{S\dagger} (\pi)     e^{-i H_{J} t_p/2} \mathcal{U}_{z} ^{S} (\pi)
$
, where $\theta = 2\pi Jt_p$. As discussed in Ref.~\onlinecite{Bian2017}, this operation is equivalent to applying only the $I_zS_z$ interaction for time $t_p$.
Likewise, we can implement $\mathcal{U}_{xx} (\theta)$ and $\mathcal{U}_{yy} (\theta)$.
In the computational basis of a two-spin system (for details see Supplementary Section \Rmnum{4}),
the CNOT gate can be realized with the sequences\cite{Vandersypen2005}
$\mathcal{U}_{\textrm{CNOT}} = \sqrt{i} \mathcal{U}_z^I(\frac{\pi}{2})  \mathcal{U}_z^S(-\frac{\pi}{2}) \mathcal{U}_x^S(\frac{\pi}{2}) \mathcal{U}_{zz} (\pi) \mathcal{U}_y^S(\frac{\pi}{2})$.
Here $I$ is the control spin, $S$ is the target spin, $\mathcal{U}_z^I(\frac{\pi}{2})$ denotes a $\pi/2$ rotation of the spin $I$ about $z$, $\mathcal{U}_z^S(-\frac{\pi}{2})$ denotes a $- \pi/2$ rotation of the spin $S$ about $z$, and so on.

\begin{figure}[http]
\centering
\includegraphics[width=1\columnwidth]{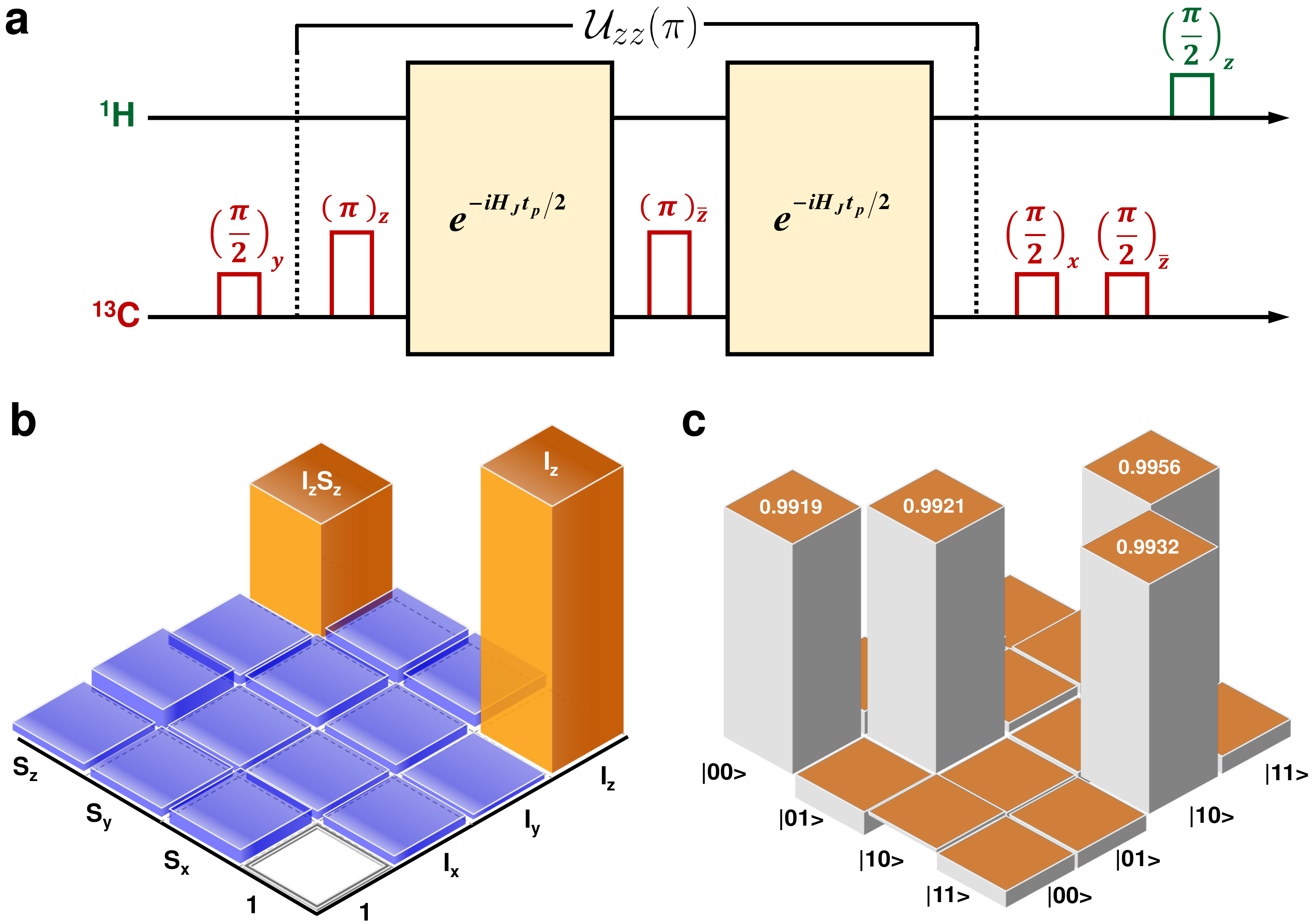}
    \caption{$\textbf{a}$, Pulse sequences for implementing CNOT gate. The $\mathcal{U}_{zz}(\pi)$ operation (see main text) is accomplished with composite pulses. The entire duration of the CNOT-gate sequence is $2.7$ ms. $\textbf{b}$, The result of state tomography after applying CNOT gate. The initial state before the CNOT gate operation is selected as the sudden state. $\textbf{c}$, The results of the CNOT gate in the computational basis. The fidelity of the CNOT gate is $0.99$.
 }
    \label{Fig3}
\end{figure}

In our experiment,
the CNOT gate is designed to flip the $^{13}$C (target spin) nuclear spin if the $^1$H (control spin) nuclear spin is in the $|$$\downarrow \rangle$ state.
Figure \ref{Fig3}a shows the pulse sequence for implementing CNOT gate in the heteronuclear two-spin system.
In order to evaluate the gate performance, standard quantum process tomography was exploited in previous work\cite{Childs2001, O'Brien2004}.
However, experimental realization of the quantum process tomography becomes resource demanding in the evaluation of the CNOT-gate performance.
We evaluate the CNOT gate using a different reconstruction technique,
in which we measure the input and output states and find the closest-fit $\mathcal{U}_{\textrm{CNOT}}$ subject to a set of constraints.
We prepare two independent initial states with fidelity above $0.98$ as the input states and measure the corresponding output states after applying the CNOT gate.
The first initial state before applying the CNOT gate is selected as the sudden state, of which the state tomography is expressed in the Pauli basis and is shown in Fig. \ref{Fig1}d.
Figure. \ref{Fig3}b shows the state tomography after the CNOT gate is applied.
Comparing the initial state (Fig. \ref{Fig1}d) with the final state (Fig. \ref{Fig3}b), it is obvious that the CNOT gate keeps $I_z$ and changes $S_z$ to $I_zS_z$, which agrees well with the theoretical calculations (Supplementary Table S$3$).
The same process was implemented with another initial state to further constrain the fit (Supplementary Section \Rmnum{4}).

Based on the state tomography results mentioned above,
we reconstruct the CNOT gate by using a numerical minimization technique to find the minimum of the function $f(\mathcal{U}_{\textrm{CNOT}})= k_1\cdot||\mathcal{B}_1||_l+k_2\cdot||\mathcal{B}_2||_l$,
where $||\cdot||_l$ denotes $l\textrm{-norm}$, $\mathcal{B}_i=\mathcal{U}_\textrm{CNOT}\rho_i\mathcal{U}_\textrm{CNOT}^{\dagger}-\rho_i^{\textrm{CNOT}}$, $k_i$ is the weighting factor of $||\mathcal{B}_i||_l$,
$\rho_i$ and $\rho_i^{\textrm{CNOT}}$ are the state tomography results before and after CNOT operation, respectively.
Although the state tomography results are more clearly presented in the Pauli basis,
they are transferred
to the computational basis for CNOT gate reconstruction (see Supplementary Section \Rmnum{4}).
Two additional constraints are added.
The first constraint is that $\mathcal{U}_{\textrm{CNOT}}$ is unitary, which means $\mathcal{U}_{\textrm{CNOT}} \mathcal{U}_{\textrm{CNOT}}^\dagger=\mathbb{1}$.
Additionally, considering the high-fidelity performance of single-spin gates,
and the coordinate orientations in our experiment, the second constraint is that
$\mathcal{U}_{\textrm{CNOT}}(1,1)\geq0$, $\mathcal{U}_{\textrm{CNOT}}(2,2)\geq0$, $\mathcal{U}_{\textrm{CNOT}}(3,4)\geq0$, and $\mathcal{U}_{\textrm{CNOT}}(4,3)\geq0$,
where $\mathcal{U}_{\textrm{CNOT}}(i,j)$ is the corresponding matrix component of the CNOT gate in the computational basis.
The form of the CNOT gate computed by finding the minimum of $f(\mathcal{U}_{\textrm{CNOT}})$ is shown in Fig. \ref{Fig3}c.
The CNOT-gate fidelity is then directly calculated to be ${{F}}=\frac{1}{4}\textrm{Tr}[\mathcal{U}^T_{\textrm{ideal}}\mathcal{U}_{\textrm{CNOT}}]=0.99$.


\section*{Discussion}
In this work, we report an experimental implementation of universal quantum control in zero-field NMR,
i.e., single-spin and two-spin control on a heteronuclear two-spin system.
Furthermore, we have evaluated the quality of the control using quantum-information-inspired randomized benchmarking and state tomography.
We have demonstrated single-spin control with an average gate fidelity of $0.9960(2)$ for $^{13}$C and two-spin control with a CNOT gate having a fidelity of $0.99$ in $^{13}$C-formic acid.
We have determined that the dominant errors for nuclear-spin control in zero-field NMR are mainly from pulse imperfections.
In addition, we have developed a temporal averaging technique to realize quantum-state tomography, which allows for full characterization of the state of the nuclear-spin system at zero magnetic field.
Although the $^{13}$C-$^1$H system is the simplest case to achieve nuclear-spin control,
our approach can be extended to a more general heteronuclear multi-spin system\cite{Bian2017}.
It is worth noting that despite scalability concerns that plague NMR ensemble quantum computing\cite{Gershenfeld1997}, zero-field NMR at the single-molecule level (detected, perhaps, with nitrogen-vacancy centers in diamond\cite{ShiF2015, Lovchinsky2016}) could take advantage of the control techniques presented here to construct a nuclear-spin quantum computer.

At zero field, as spin-spin interactions are not truncated, a multi-nuclear spin system can in principle provide more interaction types compared with the high-field case\cite{Blanchard2015}.
From this point of view, it is thus advantageous to build a quantum simulator in zero-field NMR, which can provide an efficient way to simulate physical problems,
such as Lee-Yang zeros\cite{Peng2015} and quantum magnetism\cite{Kim2010}.

Our quantum control scheme can also be applied to prepare special quantum states such as the NOON state\cite{Jones2009, Simmons2010}.
The NOON state is a quantum-mechanical many-body entangled state and is essential for quantum metrology.
However, its quantum properties may deteriorate in the presence of magnetic fluctuations\cite{Matsuzaki2011},
which can be easier to control at zero magnetic field.
Entanglement-assisted quantum metrology is promising to enhance the magnetic-field sensitivity of nuclear-spin sensors,
which provide a possibility to explore fundamental physics beyond the standard model, such as spin-axion interactions\cite{budker2014, Garcon2017}, with zero-field NMR.

Moreover, recent related work has demonstrated control of nuclear spin states at zero field with selective pulses\cite{Sjolander2017, Tayler2016} and coherent spin-decoupling\cite{Sjolander20172}.
Combining with these achievements, our work significantly extends the range of possible applications of zero-field NMR, including but not limited to fields like hetero/homonuclear decoupling, multidimensional spectroscopy, and high resolution spectroscopy.


\begin{methods}
\subsection{Sample preparation.}

${}^{13}$C-formic acid was obtained from Sigma-Aldrich.
The sample ($\sim 200$~$\mu$L) was flame-sealed under vacuum in a standard $5$~mm glass NMR tube following five freeze-pump-thaw cycles in order to remove dissolved oxygen,
which is otherwise a significant source of relaxation at zero field.

\subsection{Experimental setup.}
The oscillating magnetic field signal generated from the sample is measured by a spin-exchange-relaxation-free (SERF) atomic magnetometer, shown in Fig. \ref{Fig1}b.
The $^{87}$Rb atoms in a vapor cell are pumped with a circularly polarized laser beam propagating in the $y$ direction.
The laser frequency is tuned to the center of the buffer-gas (N$_2$) broadened and shifted $\textrm{D}1$ line.
The magnetic field is measured via optical rotation of linearly polarized probe laser light at the $\textrm{D}2$ transition propagating in the $x$ direction.
The vapor cell is resistively heated to $180$ $^\circ$C.
The atomic vapor cell is placed inside a four-layer magnetic shield (MS-1F, Twinleaf LLC),
which includes three layers of mu-metal and one innermost layer of ferrite,
which minimizes thermal Johnson noise\cite{Kornack2007}.
A set of three orthogonal coils is used to compensate the residual magnetic field to below $10^{-10}$~T.
The sensitivity of the atomic magnetometer along the $z$ axis is optimized to about $10$~$\textrm{fT}/\sqrt{\textrm{Hz}}$ for frequencies above 100~Hz.
Thus the $z$ component of the nuclear magnetization of the sample can be detected.
Three sets of mutually orthogonal low-inductance Helmholtz coils are used to apply magnetic field pulses.
\end{methods}

\subsection{Initial state preparation and readout.}
The NMR sample is polarized in a Halbach magnet ($B_p=1.8$~T), and then pneumatically shuttled down into the zero-field region within $300$~ms.
During the shuttling, a static magnetic field (guiding field, $B_g\sim 3\times 10^{-5}$~T) is applied in the $z$ direction by a solenoid wrapped around the shuttling tube.
There are two ideal cases corresponding to adiabatic and sudden transfers.
When the guiding field $B_g$ ($|(\gamma_I-\gamma_S)B_g|\gg 2\pi J$) is turned off within $10$~$\mu$s,
the state of the nuclear-spin system remains the high-field equilibrium state $\rho=\dfrac{{e}^{-H_z/k_BT}}{\textrm{Tr}({e}^{-H_z/k_BT})}$.
Here, $H_z=-B_p(\gamma_I I_z+\gamma_S S_z)$,
$k_B$ is the Boltzmann constant and $T$ is the temperature of the sample.
In the high-temperature approximation, $\rho=\frac{1}{4}(\mathds{1}+\frac{\gamma_IB_p}{k_BT}I_z+\frac{\gamma_SB_p}{k_BT}S_z)$, which is the sudden state.
Here, $\frac{\gamma_IB_p}{k_BT}\approx 1.2\times 10^{-5}$ and $\frac{\gamma_SB_p}{k_BT}\approx 3 \times 10^{-6}$.
Alternatively, when the guiding field is slowly turned off (with the characteristic time scale defined by the strength of the scalar $^{13}$C-$^1$H spin-spin coupling),
the populations at high field are converted to the populations of the zero-field eigenstates according to $\rho=\frac{(\gamma_I+\gamma_S)B_p}{8k_BT}(I_z+S_z)- \frac{(\gamma_I-\gamma_S)B_p}{4k_BT}(I_xS_x+I_yS_y)$,
which is the adiabatic state\cite{Emondts2014}.
In this experiment,
the sufficient exponential decay time to ensure adiabaticity is about $5$~s.
State tomography is implemented to measure the sudden and adiabatic states, shown in Fig. \ref{Fig1}c and d.

\begin{addendum}
\item [Acknowledgements]We thank Alexander Pines, Jiangfeng Du, Ulrich Poschinger, Dieter Suter, Ferdinand Schmidt-Kaler, Arne Wickenbrock, and Jonathan King for helpful discussions and comments, and Rom$\acute{\textrm{a}}$n Picazo Frutos for useful CNOT gate calculations.
    We would like to particularly thank Tobias Sjolander for his assistance with initial early-stage experimental efforts.
M. J. would like to acknowledge support from the China Scholarship Council (CSC) enabling his research at the Johannes Gutenberg-University Mainz.

\item[Author contributions]
M. J. designed the pulse sequences, performed experiments, analyzed the data and wrote the manuscript.
T. W. set up and characterized the SERF magnetometer, performed the measurements, implemented state tomography and wrote the manuscript.
J. W. B. designed and constructed the apparatus, devised the experimental protocol, analyzed the data and helped write manuscript.
G. R. F. designed the randomized benchmarking sequences and analyzed the data.
X. H. P. proposed the experimental concept, devised the experimental protocols, and proofread and edited the manuscript.
D. B. provided the overall management of the project, contributed to the design of the experiment and wrote the manuscript.
All authors contributed with discussions and to the final form of the manuscript.

\item[Competing Interests] The authors declare that they have no competing financial interests.

\item[Additional information]

\end{addendum}

\end{document}